\documentclass[a4paper,12pt]{article}

\usepackage{amsmath,amssymb,amscd,mathtools,array,empheq}
\usepackage[utf8]{inputenc}
\usepackage{lmodern,newtxtext,newtxmath}
\usepackage[a4paper,tmargin=3truecm,bmargin=3truecm,hmargin=2.4truecm]{geometry}
\usepackage[unicode,colorlinks=true, urlcolor=blue, citecolor=blue, linkcolor=blue]{hyperref}
\usepackage{graphicx}
\usepackage{color}
\usepackage[dvipsnames]{xcolor}
\usepackage[english]{babel}
\usepackage{bm}
\usepackage{setspace}
\usepackage{cite}
\usepackage{etoolbox}
\apptocmd{\thebibliography}{\setlength{\itemsep}{0pt}}{}{}
\usepackage[shortlabels]{enumitem}
\numberwithin{equation}{section}

\usepackage{tikz}
\usetikzlibrary{shapes,arrows,arrows.meta,matrix,decorations.markings,calc,babel,quotes,angles,math}
\makeatletter \g@addto@macro\@floatboxreset\centering \makeatother

\DeclareMathOperator{\const}{const}

\newcommand{\ie}{\textit{i.e.} }

\newcommand{\half}{\frac{1}{2}}

\newcommand{\cE}{\mathcal{E}}

\newcommand{\cO}{\mathcal{O}}

\newcommand{\bq}{\boldsymbol{q}}
\newcommand{\bp}{{\boldsymbol{p}}}
\newcommand{\br}{{\boldsymbol{r}}}
\newcommand{\bs}{{\boldsymbol{s}}}

\newcommand{\bE}{\boldsymbol{E}}
\newcommand{\bB}{\boldsymbol{B}}

\usepackage{scalerel}
\let\savewidetilde\widetilde
\def\widetilde#1{%
 \ThisStyle{\savewidetilde{\phantom{\SavedStyle#1}}%
  \setbox0=\hbox{$\SavedStyle#1$}\kern-\wd0#1}}

\tikzset{middlearrow/.style={
        decoration={markings,
            mark= at position #1 with {\arrow{>}} ,
        },
        postaction={decorate}
    }
}
\tikzmath{\lc = 0.5;}

\title{Induced motions on Carroll geometries}
 
\author{L. Marsot\footnote{mailto: marsot3@mail.sysu.edu.cn}\\[1.em]
{\normalsize School of Science, Shenzhen campus of Sun Yat-sen University,} \\
{\normalsize No. 66, Gongchang Road, Guangming District, Shenzhen, Guangdong 518107, P.R. China}}

\date{{\footnotesize (\today)}}

\begin{document}

\maketitle

\begin{abstract}
In this article, we consider some Carrollian dynamical systems as effective models on null hypersurfaces in a Lorentzian spacetime. We show that we can realize Carroll models from more usual ``relativistic'' theories. In particular, we show how ambient null geodesics imply the classical ``no Carroll motion'' and, more interestingly, we find that the ambient model of chiral fermions implies Hall motion on null hypersurfaces, in agreement with previous intrinsic Carroll results. We also show how Wigner-Souriau translations imply (apparent) Carroll motion, and how ambient particles with a non vanishing gyromagnetic ratio cannot have a Carrollian description.
\end{abstract}

\section{Introduction}

The Carroll group, originally introduced as an alternative contraction of the Poincar\'e group, where the speed of light is sent to 0 instead of infinity as for the Galilei group \cite{LevyLeblond65,SenGupta66,BacryLL68}, has attracted a lot of attention over the past decade. Carroll physics is indeed a vast topic. It can range from studying Carroll gravity, \textit{e.g.} \cite{Hartong15,BergshoeffGRRV17,HansenOOS22}, to Carroll particles/motion, \textit{e.g.} \cite{BergshoeffGL14,BoerHOSV21,MarsotZCH22,FigueroaPP23,CiambelliG23}, or use Carroll symmetries to study phenomenon in General relativity, such as on black holes \cite{Penna18,DonnayM19,FreidelJ22} or for flat holography \cite{CiambelliMPP18,DonnayFHR22}.

Now, some areas of Carroll physics remain somewhat mysterious with respect to their physical interpretation. In particular, Carroll particles or Carroll motion. This is due to the different ways one can derive Carroll physics, which do not always come with a canonical physical interpretation. Moreover, it is sometimes not clear if models for Carroll particles can actually be found in Nature. The aim of the present paper is to clarify some points regarding the physical interpretation of Carroll models, and to provide a realization for some Carroll models as effective theories on null hypersurfaces from some ``relativistic'' theories.

Broadly speaking, works on Carroll physics can be classified into 3 categories:
\begin{enumerate}[i)]
\item Firstly are the works which follow the historical definition of the Carroll group, by considering the limit $c \rightarrow 0$ of Poincaré theories. 
Examples of such works include the Carroll limit of gravitation theories, \textit{e.g.} \cite{BergshoeffGRRV17,HansenOOS22}, or the limit of dynamical models for particles/fields, see \textit{e.g.} \cite{BergshoeffGL14,HenneauxS21,BoerHOSV21,CiambelliG23,BergshoeffCFMR23}.
\item Secondly, one can work intrinsically by working directly with Carroll symmetries, without any limit/contract from Poincaré models, and derive models accordingly. See \textit{e.g.} \cite{BacryLL68,Marsot21,MarsotZCH22,FigueroaPP23} for particles, or \cite{Hartong15} for Carrollian gravity.
\item Lastly, Carroll theories can be seen as effective theories in some areas of General Relativity. This is due to the fact that Carroll geometries can be found inside Lorentzian spacetimes, as we will recall below, and this does not involve any limit or contraction of the speed of light. See for instance \cite{Penna18,DonnayM19,FreidelJ22} where Carroll physics is shown to take place on the black hole horizon, or on null infinity \cite{DuvalGH14,Herfray21}, or \cite{CiambelliMPP18,DonnayFHR22} where it is shown to be relevant for flat holography.
\end{enumerate}

The first category is convenient to work in, as the limit $c \rightarrow 0$ is usually straightforward to implement (though sometimes one has to be careful by choosing the appropriate ``electric'' or ``magnetic'' limit of the theory, just as for the Galilean limit of some theories, see \textit{e.g.} \cite{DuvalGHZ14}), however the drawback is that the physical interpretation of the resulting theory is not immediate.
Carroll theories are indeed different from Galilean theories, whose interpretation of the limit of the speed of light going to infinity is straightforward, as we know that Galilean physics is an approximation of Lorentzian physics. In other words, Galilean physics is the physics of a different realm, which is universally accessed by taking the limit $c \rightarrow \infty$.  On the other hand, the Carroll limit $c \rightarrow 0$ is not an approximation of our ``slow sort of'' world, even though it may be suitable to describe certain phenomenons. This means that the $c \rightarrow 0$ limit does not come with a canonical physical interpretation, unlike the Galilean limit.

Working intrinsically with the second category is most comfortable, because one does not leave the realm of mathematics and its rigor. The price to pay is that, of course, such derivation for Carrollian models does not give any canonical physical interpretation for the parameters of the models. The models derived in such a way need to then be compared on physical examples to be interpreted. For instance, as noted in \cite{MarsotZCH22,FigueroaPP23}, some example of Carrollian models can be fit on physical examples where the (conserved) quantity associated to Carroll time translations is identified as a mass, an energy, or even an electric charge. Moreover, it may happen that some mathematical models obtained in this way, while mathematically correct, have no physical realization (the same is also true for models obtained in the first category).

The third category is the most fruitful with respect to physical interpretation. In this category, Carroll physics is seen as an effective description of a bulk theory which we understand (or at least interpret) well, which means that the interpretation of the parameters of the Carroll theory is deduced from that of the bulk description\footnote{The case can be made that some theories in the first category may also be seen as an effective description of some physical system. However due to the contraction process, the link between parameters is not necessarily clear.}.
For instance, it is now well-known that Carroll geometries are naturally present on null hypersurfaces in Lorentzian spacetimes \cite{Morand18,CiambelliLMP19}, and as such, one realization of Carroll physics is as an effective description of such hypersurfaces\footnote{This is not the only possible physical realization of Carrollian models: Carroll physics may also be relevant in the context of condensed matter, or for fractons.}. Such Carrollian effective descriptions can also be compared to models obtained in the first two categories to give them a physical realization.

Note that the border between the first and third category may be blurry at times, as the process of taking the limit of $c$ to 0 may be formally very similar to the process of restricting the geometry to some submanifold. One may also take the limit of parameters other than the speed of light, see \textit{e.g.} \cite{deBoerHOSV23}.

Carrollian dynamical systems, \textit{e.g.} Carroll particles, are usually studied in the first and second categories. The purpose of this paper is to investigate Carrollian dynamical systems from the point of view of the third category. Hopefully by doing so, it will be possible to identify models obtained as a contraction or intrinsically, which do not yet have any  physical realization. For instance, recently some Carroll fluids were identified in this manner \cite{BagchiKS23}.

Note that in this paper, we are ultimately interested in motion induced on null hypersurfaces which are Carroll geometries. At the level of symmetries, the Carroll boosts are not necessarily induced from the symmetries of the bulk theory\footnote{If one removes Carroll boosts, the symmetries then span the Aristotle group. However, it has been shown \cite{MarsotZCH22} that (free) Aristotle-invariant models display accidental symmetries and are, anyway, invariant under the Carroll group.}.

In section~\ref{s:naive}, we will review the most basic dynamical model, null geodesics. We will show how null geodesics induce trivial motion on null hypersurfaces in a Minkowskian spacetime, which leads to the well-known ``no-motion'' Carrollian dynamical model. We will use this sort of naive introduction to effective models on null hypersurfaces as a way to reflect on the physical parameters which are induced in the effective model.
In the next section~\ref{s:chiral fermions}, we consider the (Lorentzian) model of a charged massless particle, which would be the semiclassical counterpart of a massless chiral fermion, and its effective description on Carroll null hypersurfaces. We also use this section to introduce the notations and formalism used in the rest of the paper.
Later, in section~\ref{s:carroll wigner}, we talk about the relevance of Wigner-Souriau translations for inducing Carroll models on null hypersurfaces, as they imply that the induced motions are observer dependent (for an observer in the bulk).
In the last section~\ref{s:gfactor}, we generalize the section~\ref{s:chiral fermions} (and \ref{s:carroll wigner}) by considering particles with a non-zero gyromagnetic ratio $g$.

\section{The effective Carrollian description of null geodesics}
\label{s:naive}

Carroll dynamics naturally take place on Carroll structures. Such structures appear commonly in General Relativity, as null hypersurface in a Lorentzian spacetime naturally display Carroll geometries \cite{Morand18,CiambelliLMP19}. Distinguished examples being black hole horizons, and null infinity.

Null hypersurfaces are naturally the realm of massless particles (or tachyons). The equations of motion for such massless particles in flat spacetime are the null geodesic equations and are plainly given by the set of first order equations,
\begin{subequations}
\label{geodesics}
\begin{align}
\dot{X}^\mu & = P^\mu\,, \label{geodesics_dx} \\
\dot{P}^\mu & = 0\,, \label{geodesics_dp}
\end{align}
\end{subequations}
where the dot denotes the normal derivative on $X$ and the covariant derivative on $P$, with respect to some arbitrary affine parameter. 

Let us pick coordinates $(x, y, z, t)$ such that the momentum of the massless particles we consider reads
\begin{equation}
\label{mom_outoing}
(P^\mu) = \frac{\cE}{c} \left(\begin{array}{c}
0 \\
0 \\
\pm 1 \\
1
\end{array}\right)
\end{equation}
such that particles with $P^z > 0$ are called outgoing and those with $P^z < 0$ are called ingoing.
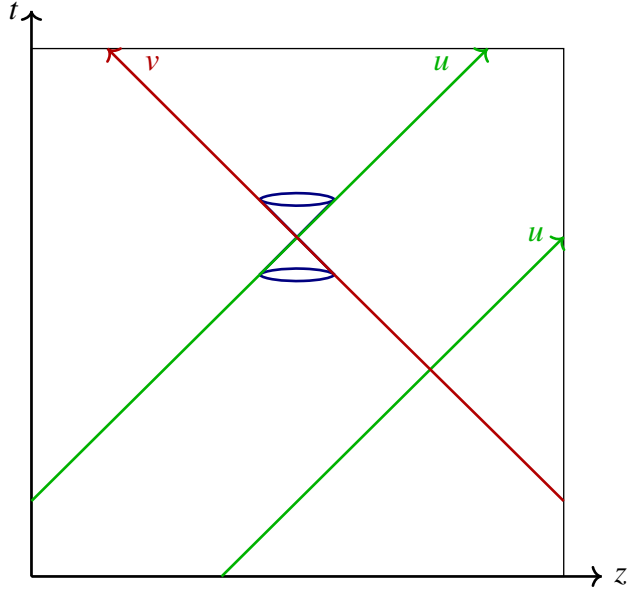
\begin{figure}[ht]
\begin{tikzpicture}[line width=1pt, every node/.style={transform shape}]
  \draw [line width=0.5pt] (0,0) coordinate (bottomleft) -- (0,7) coordinate (topleft) -- (7,7) coordinate (topright) -- (7,0) coordinate (bottomright) -- (bottomleft);
  \coordinate (bottomh) at (2.5,0);
  \draw [line width=1pt,middlearrow={1}] (bottomleft) -- node[pos=1,left]{$t$} (0,7.5);
  \draw [line width=1pt,middlearrow={1}] (bottomleft) -- node[pos=1,right]{$z$} (7.5,0);

  \draw [black!50!blue](3,4) arc (170:10:{1*\lc} and {0.2*\lc}) coordinate[pos=0] (a);
  \draw [black!50!blue](a) arc (-170:-10:{1*\lc} and {0.2*\lc}) coordinate (b);
  \draw [black!50!blue](a) -- ([yshift={2cm*\lc}]$(b)$) coordinate (c);
  \draw [black!50!blue](b) -- ([yshift={2cm*\lc}]$(a)$) coordinate (d);
  \draw [black!50!blue](d) arc (170:10:{1*\lc} and {0.2*\lc});
  \draw [black!50!blue](d) arc (-170:-10:{1*\lc} and {0.2*\lc});

  \draw [black!30!green,middlearrow={1}] (2.5,0) to[out=45,in=-135] node[pos=0.93,yshift=0.4cm]{$u$} (7,4.5);
  \draw [black!30!green,middlearrow={1}] (0,1) to[out=45,in=-135]  (a) -- node[pos=0.8,yshift=0.4cm]{$u$} ([xshift=3cm,yshift=3cm]a);

  \draw [black!30!red,middlearrow={1}] (7,1) to[out=135,in=-45] (b) --  node[pos=0.80,yshift=0.4cm]{$v$} (1,7);

\end{tikzpicture}
\caption{ (1+1 dimensional) Spacetime diagram. The light-cone is in blue. There are two kind of trajectories for photons, either \emph{outgoing} (in the sense that $p_z > 0$) depicted in green, or \emph{ingoing} ($p_z < 0$) depicted in red.}
\label{f:spacetime}
\end{figure}

Now, the direction of ingoing and outgoing null geodesics define new coordinates that are well suited for the study of null trajectories. These null coordinates $u$ and $v$ an be introduced such that the Minkowski flat metric takes the form $ds^2 = dx^2 + dy^2 + 4 du dv$, which is accomplished by the coordinate change,
\begin{equation}
\label{null_coordinates}
u = \frac{z + ct}{2}\,, \quad
v = \frac{z - ct}{2} \,.
\end{equation}

The next step is to specify the geodesic equation of motion \eqref{geodesics} for a massless particle that is, for instance, \emph{outgoing}, \textit{i.e.} with momentum \eqref{mom_outoing}, where we choose the positive sign, or when using null coordinates, $P = \cE/c \, \partial_u$. We readily see that the equations of motion \eqref{geodesics_dx} imply $\dot{u} = 1$ and $\dot{v} = 0$ (after choosing the appropriate affine parameter). This means that outgoing massless particles have their trajectories along a $v = \const$ hypersurface with $u$ a natural evolution parameter along the trajectory. Conversely, a massless particle that is ingoing would have $P = - \cE/c \partial_v$ and follow $u = \const$ hypersurfaces, with $v$ the natural evolution parameter. The situation is depicted in figure~\ref{f:spacetime}.

A (2+1 dimensional) Carroll structure $(\Sigma_{v}, g_\Sigma, \xi)$ \cite{Henneaux79,DuvalGHZ14} can be obtained as a null hypersurface $\Sigma_{v}$ defined as the resulting geometry from taking a $v = \const$ slice of $3+1$-dimensional spacetime. In that case, one gets the degenerate metric and a vector field on $\Sigma_{v}$,
\begin{equation}
g_\Sigma = dx^2 + dy^2, \quad  \xi = \partial_u.
\end{equation}

The equations of motion \eqref{geodesics_dx} for an ``outgoing'' photon trivially mean that it stays on the hypersurface $\Sigma_{v}$ defined by $v = \const$. The equations \eqref{geodesics_dx} for our outgoing photon then yield, 
\begin{equation}
\label{induced_carroll_eom}
\frac{d x}{du} = 0\,, \qquad \frac{d y}{du} = 0\,,
\end{equation}
where we used $u$ as the evolution parameter, since $\dot{u} = 1$. These equations of motion are the effective equations of motion induced on the Carroll structure $(\Sigma_{v}, g_\Sigma, \xi)$. They are of course trivial: the particle does not move on the null hypersurface. While the result is trivial, this effective description recovers, from the point of view of the 3rd category defined in the introduction, the well-known results from \cite{DuvalGHZ14,BergshoeffGL14} that the simplest Carroll model of a particle does not move.

All of the above is technically trivial, but is a good way to understand why the ``no-motion'' statement for Carroll particles is so natural. Indeed, null hypersurfaces defined by $v = \const$ are Carroll structures. Such a structure is described by 2 spatial coordinates $x$ and $y$, and another coordinate which is ``along'' the null (w.r.t bulk) direction, $u$. The coordinate $u$ is usually called the ``Carroll time''. Hence, in this context of null hypersurfaces, the intrinsic statement that Carroll particles do not move simply mean that particles move in the null direction that defines the hypersurface, \ie that it follows a null geodesic. The massless particle has all its momentum directed toward the null direction, and has no momentum to ``spare'' to deviate in the other, spatial, directions. In this context, the statement of the Red Queen, that \textit{it takes all the running you can do, to keep in the same place} is particularly apt to describe Carrollian physics.

Carrollian models (derived intrinsically or through the Carrollian limit) which do feature motion could then only be deduced, in the context of induced Carroll structures on null hypersurfaces in a Lorentzian spacetime, from ambient models which predict derivations to null geodesics. 

\medskip

Let us now talk about conserved physical quantities and physical interpretation. In the bulk, if the model is invariant under space and time translations, the momentum $P$ of the particle is conserved. Note that, since the conserved quantities are of the form $\Psi = P_\mu K^\mu$ for a vector field $K$ representing the symmetries, when using light cone coordinates  \eqref{null_coordinates}, the symmetry $K^u \partial_u$ (resp. $K^v \partial_v$) will generate the conserved quantity $P^v$ (resp. $P^u$).

Furthermore, to stay on a null hypersurface spanned by one of the null directions, the momentum of the massless particle necessarily vanishes in the other null direction. For instance, if $\partial_u$ (resp. $\partial_v$) generates Carroll time on the null hypersurface, then $P^v$ (resp. $P^u$) vanishes. 

The above two paragraphs mean that the conserved quantity induced on the Carroll geometry, \textit{i.e.} the one associated to Carroll time translations, necessarily vanishes\footnote{Note that the naming convention in the literature for the conserved quantity associated to Carroll time translations can be somewhat confusing. It may be called an energy, a mass, or even an electric charge. There is of course no canonical naming convention, the name depends on the physical application of the Carroll group. Calling this parameter ``energy'' seems more popular in works involving the limit $c \rightarrow 0$,  \textit{e.g.} in \cite{BergshoeffGL14}, while calling it a ``mass'' seems more common in works involving null hypersurfaces, see \textit{e.g.} \cite{DuvalGHZ14}.}. The other null component of the ambient momentum, which encodes the energy of the particle for the model of null geodesics, does not vanish, but it is associated with a symmetry that is not induced on the Carroll geometry. Given the expression for the momentum \eqref{mom_outoing}, this means that, in the context of Carroll motions induced on a null hypersurface in a Lorentzian spacetime from a spinless and massless particle, the data of the energy of the massless particle is not induced on the Carroll geometry and is thus effectively ``lost''. This also means that in our context of motion induced on a null hypersurface, the conserved quantity associated to Carroll time translations is not related to the energy of the ambient particle.

Now, there exist many Carroll models found from coadjoint orbit methods in \cite{Marsot21,MarsotZCH22,FigueroaPP23}, some of which allow motion, and some of which even correspond to ``massive'' models where the conserved quantity along Carroll time translations does not vanish. It is then natural to wonder whether these non-trivial intrinsic Carroll models can be realized on null hypersurfaces from more elaborate bulk models than just massless particles following null geodesics.

\section{Massless chiral fermions}
\label{s:chiral fermions}

The particles we choose to study in this section are massless, have spin, and as we want to couple them to electromagnetism, we allow them to carry an electric charge, $e$. Such particles (which do arise in condensed matter physics \cite{Venemaetal16}) would then be the semiclassical counterparts of \emph{massless chiral fermions} \cite{SonY12,StephanovY12,ZhangH14}.

From now on, we use units such that $c = 1$.

The trajectory of such particles a Lorentzian spacetime has been described in \cite{DuvalH14} by the following set of equations,
\begin{subequations}
\label{eom_g0}
\begin{align}
\dot{X}^\mu & = P^\mu + 2 \frac{S^\mu{}_\nu F^\nu{}_\rho P^\rho}{S \cdot F}, \label{g0_dx} \\
\dot{P}^\mu & = -e F^\mu{}_\nu \dot{X}^\nu, \label{g0_dp} \\
\dot{S}^{\mu\nu} & = P^\mu \dot{X}^\nu - P^\nu \dot{X}^\mu.
\end{align}
\end{subequations}
where $S$ is a skew-symmetric 2-tensor representing the spin state of the particle, $F$ is the electromagnetic 2-tensor, $S \cdot F := S^{\rho\sigma}F_{\rho\sigma}$, and where the dot over $P$ and $S$ denote the covariant derivative, while it denotes the usual derivative on $X$.

These equations are found in \cite{DuvalH14} as spanning the kernel of the presymplectic 2-form \cite{Souriau70},
\begin{equation}
\label{sigma_chiral}
\sigma = -d P_\mu \wedge dX^\mu - \frac{1}{2 s^2} dS^\mu{}_\lambda \wedge S^\lambda{}_\rho dS^\rho{}_\mu + \frac{e}{2} F_{\mu\nu} dX^\mu \wedge dX^\nu,
\end{equation}
as well as obeying the following \emph{supplementary conditions} (which are somewhat akin to equations of state for the particle),
\begin{equation}
\label{supplementary conditions}
P^2 = 0, \quad \half S^{\mu\nu}S_{\mu\nu} = s^2, \quad S^\mu{}_\nu P^\nu = 0,
\end{equation}
with $s$ the scalar, or longitudinal spin of the particle.

The last condition, that $P$ is in the kernel of $S$ is called the Tulczyjew condition \cite{Tulczyjew59}. Note that the Tulczyjew condition, together with the fact that $S$ is skew-symmetric and spacetime is of even dimension, implies the existence of another vector $J$, not parallel to $P$, such that $S^\mu{}_\nu J^\nu = 0$.  These relations imply that we can decompose the spin tensor as,
\begin{equation}
\label{decomp_S}
S_{\mu\nu} := s \, \epsilon_{\mu\nu\rho\sigma} \frac{P^\rho J^\sigma}{P\cdot J},
\end{equation}
and one can then choose $J$ null such that
\begin{equation}
\label{conds_ij}
P^2 = J^2 = 0, \quad P_\mu J^\mu = -1.
\end{equation}

The equations of motion \eqref{eom_g0} can also be obtained by starting from the well known Mathisson-Papapetrou-Dixon (MPD) equations coupled to electromagnetism (with vanishing gyromagnetic ratio, \ie for $g = 0$), \cite{Mathisson37,Papapetrou51,Dixon70,Souriau74},
\begin{subequations}
\label{mpd}
\begin{align}
\dot{P}^\mu & = -e F^\mu{}_\nu \dot{X}^\nu
\\
\dot{S}^{\mu\nu} & = P^\mu \dot{X}^\nu - P^\nu \dot{X}^\mu \label{mpd_s}
\end{align}
\end{subequations}
and asking the supplementary conditions \eqref{supplementary conditions} to hold.

Indeed, differentiating $S^\mu{}_\nu J^\nu = 0$ and replacing $\dot{S}^\mu{}_\nu$ by its expression from the MPD equations, one finds
\begin{equation}
\label{xdot_dep_j}
\dot{X}^\mu = P^\mu + \frac{1}{P_\rho J^\rho}S^{\mu}{}_\nu \dot{J}^\nu
\end{equation}
after choosing a suitable worldline parameter such that $\dot{X}^\mu J_\mu = P^\mu J_\mu$. Then, plugging the decomposition \eqref{decomp_S} and the velocity \eqref{xdot_dep_j} into \eqref{mpd_s}, one can find that $S^\mu{}_\nu F^\nu{}_\lambda \dot{X}^\lambda = 0$ which, together with \eqref{xdot_dep_j} implies,
\begin{equation}
\label{anomalous velocity equality}
\frac{1}{P_\rho J^\rho}S^{\mu}{}_\nu \dot{J}^\nu = \frac{2}{S \cdot F} S^\mu{}_\nu F^\nu{}_\rho P^\rho,
\end{equation}
thus recovering \eqref{g0_dx}\footnote{More directly, one can just differentiate the Tulczyjew condition to obtain these equations of motion. However, this method has the advantage to show that the velocity-momentum relation is not always of the form \eqref{g0_dx} which becomes degenerate in the limit of a vanishing field. In some literature, \textit{e.g.} \cite{HarteO22}, this kind of derivation is usually done with a linear combination of both $J$ and $P$ as we will see later. However the computation is formally identical to using $J$ alone as we do here. An almost identical computation shows that one can recover the ``Souriau--Saturnini'' equations \cite{Saturnini76} for a massless particle with spin in a gravitational field from, \emph{e.g.} the equations found in \cite{HarteO22}. \label{fn_ss_equiv_marius}}.

With this point of view, while the existence of the vector $J$ is clear, its interpretation is not. In the massive case, the vector $J$ is straightforward to interpret, as it is proportional to the Pauli-Lubanski vector. However, in the massless case, the Pauli-Lubanski vector is instead proportional to $P$. This lack of interpretation for $J$ makes the form of the equation \eqref{xdot_dep_j} unacceptable. This is what justifies replacing the $J$ dependency using the expression \eqref{anomalous velocity equality}. 

Note that this substitution is a somewhat tricky operation: it makes the resulting equations only valid for non vanishing $S\cdot F$, and working with the equations using numerical methods or perturbation theory becomes subtle due to the ratio of what may be two small quantities. If one can identify the vector $J$, such as in, \textit{e.g.}, \cite{OanceaK22} (by being part of the vector field representing the observer measuring the worldline), one should probably keep the explicit dependency on $J$ (or on the observer), which is the method we will adopt in the following sections.

Using usual Minkowski coordinates, the 4-momentum of a photon is $P = \cE (\bp, 1)$ and the supplementary conditions \eqref{conds_ij} imply that $J = (\bq, - 1/\cE)$ with $\vert \bq \vert = 1$, such that $\bp \cdot \bq = 0$. Now, with these notations, the spin tensor can be decomposed as by $S_{ij} = \epsilon_{ijk} s^k$ and $S_{i4} = \epsilon_{ijk} p^j s^k$, with
\begin{equation}
\bs = s(\bp + \bq)
\end{equation}

Note that the Spin-EM coupling term is $S \cdot F$, which becomes 
\begin{equation}
S\cdot F = 2 \bs \cdot \left(\bB - \bp \times \bE\right)
\end{equation}

\medskip

It was shown in \cite{DuvalH14} that, upon choosing a $3+1$ splitting such that $X = (\br, t)$, the first two equations \eqref{g0_dx} and \eqref{g0_dp} become,
\begin{subequations}
\label{eom_chiral}
\begin{align}
\dot{\br} & = s \frac{\bB - \bp \times \bE}{\bs \cdot (\bB - \bp \times \bE)}, \label{g0_rdot}\\
\dot{t} & = s \frac{\bp \cdot \bB}{\bs \cdot (\bB - \bp \times \bE)}, \label{g0_tdot} \\
\dot{\bp} & = s \frac{e \bE \cdot \bB}{\bs \cdot (\bB - \bp \times \bE)} \bp, \label{g0_pdot1} \\
\dot{p} & = s \frac{e \bE \cdot \bB}{\bs \cdot (\bB - \bp \times \bE)} \label{g0_pdot2}
 \end{align}
\end{subequations}

These equations, as already noticed in \cite{DuvalH14}, are very peculiar due to the vanishing of the linear momentum as a linear dependence on the right hand side of \eqref{g0_rdot}. 

Now, these equations are valid for massless charged fermions in flat $3+1$ spacetime. Moreover, such spacetimes can be foliated by null hypersurfaces, which are Carroll structures. Hence, let us consider again light-cone coordinates $(x, y, u, v)$, see \eqref{null_coordinates}. Just as in the previous section, we will look at the induced motion on a null hypersurface/Carroll structure $\Sigma_v$ defined by $v = \const$.

The first point to look at when inducing motion on $\Sigma_v$ is that the particle actually stays $\Sigma_v$. In our case, this means that the particle's equations of motion need to imply the vanishing of
\begin{equation}
\label{g0_vdot}
\dot{v} = \half \left(\dot{z} - \dot{t}\right) =  - s \frac{\bp \cdot \bB - B_z + p_x E_y - p_y E_x}{2 \bs \cdot (\bB - \bp \times \bE)}.
\end{equation}
Moreover, since we would like the electric and magnetic fields to be defined on the null hypersurfaces $\Sigma_v$, we will also require that
\begin{equation}
\label{field_induced_conditions}
E_z = 0, \quad B_x = 0, \quad B_y = 0.
\end{equation}
The above conditions together with the vanishing of $\dot{v}$ \eqref{g0_vdot} then implies that the momentum should be such that $p_x = p_y = 0$ and $p_z = + 1$. This corresponds to the momentum of the photon being the same as in the previous section, see \eqref{mom_outoing}, \ie $P = \partial_u$, which is the ``direction'' of Carroll time on $\Sigma_v$.

Note that the conditions \eqref{field_induced_conditions} on the electromagnetic fields readily imply that the electric and magnetic fields are orthogonal to each other, which immediately implies that the momentum of the particle is constant per the equations of motion \eqref{g0_pdot1}--\eqref{g0_pdot2}.

Finally, assuming that $\half S \cdot F \neq 0$, i.e. $s_z B_z \neq \left(\bs \times \bE\right)_z$, one then finds that the equations of motion \eqref{eom_chiral} reduce to,
\begin{equation}
\frac{d x^i}{du}  = \epsilon^{ij} \frac{E_j}{B_z}, \qquad
\frac{dv}{du} = 0, \qquad
\frac{dp^i}{du} = 0,
\label{Hallforchiral}
\end{equation}
where $i, j \in \{x, y\}$.

Those equations do not depend on $J$ or $\bq$, which is a good property given that these vectors are hard to interpret for massless particles. 

These equations of motion are exactly those obtained intrinsically for charged massless Carroll particles in an electromagnetic field in \cite{MarsotZCH22}, which are particularly interesting given that those equations are the same as for the Hall law.

Carroll dynamics have long been assumed to be trivial, like the model show in the previous section where particles do not move. However, in the past year, some non trivial (intrinsic) Carroll models have been developed, and we have just seen in this section how to recover one such model from an ambient perspective.

\section{Observer dependence of induced Carroll motion}
\label{s:carroll wigner}

In section \ref{s:naive}, we saw that it is natural to work with massless particles. While we considered photons as usually is done in GR, \textit{i.e.} as massless and spinless particles following null geodesics, photons are massless particles \emph{with spin}. The key point for this section is that it has been shown that the trajectory of such particles (massless with spin) is not well-defined, unlike for their spinless counterpart, but is observer dependent, due to \emph{Wigner-(Souriau) translations} \cite{Souriau70,StoneDZ14,DuvalH14,DuvalEHZ14,StoneDZ15}. 

We will use the same formalism to describe photons with spin here as in the previous section for chiral fermions.

The observer dependency of the trajectory of a free massless particle with spin manifests itself by that the trajectory is only determined up to a vector $Z$ orthogonal to the momentum \cite{Souriau70},
\begin{equation}
\label{condition_ws}
P \cdot Z = 0,
\end{equation}
which transforms the trajectory data $(X, P, J)$ as \cite[\S 14]{Souriau70},
\begin{subequations}
\label{transf_z}
\begin{align}
X^\mu & \mapsto X^\mu + s Z^\mu, \label{ws_x}\\
P_\mu & \mapsto P_\mu, \\
J_\mu & \mapsto J_\mu + \epsilon_{\mu\nu\lambda\rho} Z^\nu P^\lambda J^\rho  + \half Z^2 P_\mu,
\end{align}
\end{subequations}
or $S_{\mu\nu} \mapsto S_{\mu\nu} + 2 s Z_{[\mu} P_{\nu]}$.

The orbit of the motion of a massless particle with spin by the transformation \eqref{ws_x}, given the condition \eqref{condition_ws}, describes a wavefront \cite{Souriau70}, \ie it looks like a 2 dimensional plane moving in the direction of $P$.

Quite interestingly, the Carroll hypersurface defined by the direction of propagation of such a dynamical system is then the wavefront itself, in that the ambient direction of propagation becomes Carrollian time, and the 2-dimensional plane of the wavefront is the Carrollian ``space''.

Now, instead of working with the null vector $J$ as in the previous section, it is sometimes convenient to work with the velocity field $t$ of a timelike observer who sees the photon with energy $E$. To satisfy $t^2 = -1$,  $t^\mu P_\mu = -\cE$, and that $t$ is written in the base $(P, J)$, the vector field should be, 
\begin{equation}
\label{t geq0}
t = \frac{1}{2\cE} P + \cE \, J.
\end{equation}

We can then write the spin 2-tensor $S_{\mu\nu}$ using this new vector $t$ instead of $J$ as, 
\begin{equation}
S_{\mu\nu} = \frac{s}{P\cdot t} \epsilon_{\mu\nu\rho\sigma} P^{\rho} t^{\sigma},
\end{equation}
which then obeys $S_{\mu\nu} t^\nu = S_{\mu\nu} P^\nu = 0$. 

The Wigner-Souriau transformations \eqref{transf_z} then imply that $t$ also transforms by $Z$,
\begin{equation}
t_\mu \mapsto t_\mu + \epsilon_{\mu\nu\rho\sigma} Z^\nu P^\rho t^\sigma + \frac{E}{2} Z^2 P_\mu.
\end{equation}
Note that, though the velocity of the observer changes under a Wigner-Souriau translation, the observer still sees the photon with the same energy, as $P \cdot t$ is invariant. Likewise, the relations $S_{\mu\nu} t^\nu = S_{\mu\nu} P^\nu = 0$ also stay invariant under this shift.

Using the ``Souriau-Duval'' convention $(X, P, J)$ is fully equivalent to using the convention $(X, P, t)$ on a technical standpoint. The latter has however the advantage that it is easier to interpret the field $t$ of an observer than a somewhat obscure vector field $J$ for massless particles, and give it proper initial conditions\footnote{This can particularly be seen when considering massless particles in a gravitational field. In footnote \ref{fn_ss_equiv_marius}, we mentioned how the theories of \cite{Saturnini76} and \cite{HarteO22} are equivalent (provided some quantity $R(S)(S) \neq 0$ does not vanish), the former using the $(X, P, J)$ conventions, and the latter the $(X, P, t)$ conventions. The latter work is however clearer, in particular due to the ease of interpretation of initial conditions.}.

\bigskip

By choosing an observer $t$ (alternatively a null vector $J$ for a given $P$), we select a specific worldline for the massless particle. The equations of motion for this worldline is obtained as follows.

In an almost identical derivation to \eqref{xdot_dep_j}, the relation $S^\mu{}_\nu t^\nu = 0$, which is usually called the Corinaldesi-Papapetrou spin supplementary condition \cite{CorinaldesiP51}, leads to a relation between the velocity and the momentum of the particle, see \textit{e.g.} \cite{CostaNatario14},
\begin{equation}
\label{observer_xdot}
\dot{X}^\mu = P^\mu - \frac{S^{\mu\nu}}{(P \cdot t)} \dot{X}^\lambda \nabla_\lambda t_\nu.
\end{equation}

This momentum-velocity relation implies that an accelerated observer sees the photon with an anomalous velocity, provided the acceleration is not in the kernel of $S$.

This equation of motion has alternatively been derived using a WKB expansion of electromagnetic wave packets in \cite{OanceaJDRPA20,HarteO22}. In their framework, the motion of a massless particle with spin seen by an observer $t$ in flat spacetime follows the equations of motion,
\begin{subequations}
\begin{align}
\dot{X}^\mu & = P^\mu - s (F_{xp})^\mu{}_\nu P^\nu + \cO((1/\cE)^2) \label{xdot_marius} \\
\dot{P}^\mu & = 0
\end{align}
\end{subequations}
where $F$ is a Berry curvature term arising due to the observer $t$,
\begin{equation}
(F_{xp})^\mu{}_\nu = - \frac{\epsilon^{\mu\lambda\rho\sigma} P_\rho t_\sigma}{(P \cdot t)^2} \nabla_\nu t_\lambda.
\end{equation}

The two equations of motion \eqref{observer_xdot} and \eqref{xdot_marius} then coincide up to order in $1/E$.

In fact, one can show the equivalence of the two formalisms at the symplectic level, since the spin contribution appearing in the 2-form \eqref{sigma_chiral} can be transformed, using the decomposition of the spin tensor \eqref{decomp_S}, as
\begin{equation}
\frac{1}{2s^2} dS^\mu{}_\lambda \wedge S^\lambda{}_\rho dS^\rho{}_\mu = - \frac{S^{\mu\nu}}{4E^2} dP_\mu \wedge dP_\nu + \frac{S^{\mu\nu} \nabla_\nu t_\lambda}{2 E} dP_\mu \wedge dX^\lambda,
\end{equation}
which are the two Berry curvature contributions (in flat space) to the symplectic form found in \cite{OanceaJDRPA20} due to the observer ($F_{pp}$ and $F_{xp}$ in their notations).

\bigskip

To understand how this observer dependency affects the induced motion on a Carroll hypersurfaces, consider again an outgoing photon in light cone coordinates $(x, y, u, v)$, \textit{i.e.} with momentum
\begin{equation}
P = \cE \, \partial_u,
\end{equation}

The velocity field of the observer should be such that $t^2 = -1$ and $P \cdot t = - \cE$ which implies,
\begin{equation}
(t_\mu) = (t_x, t_y, -1, 1+t_x^2 + t_y^2),
\end{equation}
where $t_x, t_y$ are a priori functions on the ambient manifold.

The $v$-component of the momentum-velocity relation \eqref{observer_xdot} is then,
\begin{equation}
\label{obs_vdot}
\dot{v} = \frac{s}{\cE} \left(t_x \partial_u t_y - t_y \partial_u t_x\right),
\end{equation}
whose vanishing (so that the motion of the particle stays on $\Sigma_v$) implies $t_x \partial_u t_y = t_y \partial_u t_x$. This implies in turn for the apparent spatial velocity on $\Sigma_v$,
\begin{subequations}
\begin{align}
\dot{x} & = - \frac{s}{\cE} \left(1 - t_x^2 - t_y^2\right) \partial_u t_y \\
\dot{y} & = \frac{s}{\cE} \left(1 - t_x^2 - t_y^2\right) \partial_u t_x
\end{align}
\end{subequations}
Note that we have $\dot{u} = 1$.

If we define $Q = 1 - t_x^2 - t_y^2$, the equations of motion can be put in the form,
\begin{equation}
\label{g0_eom_spatial}
\dot{X^i} = - \frac{s Q}{\cE} \epsilon^{ij} \partial_u t_i, \quad i \in \{x, y\}.
\end{equation}

Likewise, the condition \eqref{obs_vdot} can be put in the form $t_i \epsilon^{ij} \partial_u t_j = 0$. This shows that the ``spatial'' (in reference to the Carroll spatial plane $(x,y)$) acceleration of the observer (with respect to the ``Carroll time'' $u$) should be parallel to its velocity $\partial_u t_i \propto t_i$, which in turn implies $\dot{X}^i \propto \epsilon^{ij}t_j$ from \eqref{g0_eom_spatial}. The anomalous velocity induced on a Carrollian geometry as seen by an accelerated observer is thus orthogonal to the spatial velocity of the observer.

This observer effect, or Wigner-Souriau translations, are thus another way to induce (apparent) motion on a Carroll structure. However since this effect is not intrinsic to the particle, but instead depends on an observer in the ambient spacetime, it seems difficult to link the equations of motion \eqref{g0_eom_spatial} to an intrinsic Carroll model.

\section{Massless chiral fermions with gyromagnetic ratio \texorpdfstring{$g\neq0$}{g!=0}}
\label{s:gfactor}

In section \ref{s:chiral fermions}, we worked out the equations of motion of a massless particle of charge $q$ and vanishing gyromagnetic ratio $g = 0$. A natural question is then what happens for particles with $g \neq 0$.

The model for massless chiral fermions with $g \neq 0$ is the same as in the section \ref{s:chiral fermions} for $g = 0$, with the exception of the relation \cite{Duval76,DuvalH14},
\begin{equation}
\label{p2 gneq0}
P^2 = - \frac{eg}{2} S \cdot F 
\end{equation}
see also \cite{OanceaK22} for $g = 2$. Hence, it appears as if these particles have an effective mass.

The conserved energy of such a particle then yields the dispersion relation \cite{DuvalH14},
\begin{equation}
\label{dispersion}
\cE = \sqrt{\vert \bp \vert^2 - \frac{eg}{2} S\cdot F}
\end{equation}
such that $P = (\bp, \cE)$.

The momentum $P$ can be expressed as, 
\begin{equation}
\label{def p g}
P = I + \frac{eg}{4} \left(S \cdot F\right) J,
\end{equation}
with $I$ and $J$ null and $I \cdot J = -1$, where the null vector $I$ corresponds to the momentum $P$ in the limit of vanishing gyromagnetic ratio.

\medskip

Now, we wish to switch to the conventions $(X, P, t)$ to interpret more easily the resulting equations of motion. To find the expression of the observer $t$, we again require that $t^2 = -1$ and $P\cdot t = -\cE$. Moreover, we want to recover the expression \eqref{t geq0} of $t$ in the limit where the electromagnetic field vanishes, $F \rightarrow 0$. This leads us to,
\begin{equation}
\label{t gneq0}
t = \frac{\cE - \sqrt{\vert \bp \vert^2 -eg S\cdot F}}{eg S\cdot F/2} P + \sqrt{\vert \bp \vert^2 -eg S\cdot F} \, J,
\end{equation}
which can be expanded in the weak field limit as,
\begin{equation}
t \approx \frac{1}{2 \cE} P + \frac{eg}{16 \cE^3} S \cdot F \, P + \cE J - \frac{eg}{4\cE} S\cdot F \, J,
\end{equation}
where $\approx$ is an equality up to, and including, linear terms in the spin (more rigorously, linear terms in $e S\cdot F/\cE^2$).

Note that we have
\begin{equation}
S_{\mu\nu} = - \epsilon_{\mu\nu\rho\sigma} I^\rho J^\sigma = \frac{\epsilon_{\mu\nu\rho\sigma}}{\sqrt{P^2 + (P\cdot t)^2}}P^\rho t^\sigma 
\end{equation}

A similar computation from \cite{OanceaK22}, but without neglecting higher than linear spin terms, and for a general $g$, yields the equations of motion,
\begin{subequations}
\label{eom_gneq0}
\begin{align}
\dot{X}^\mu & = P^\mu + \frac{2 (P \cdot t) S^\mu{}_\nu}{2(P\cdot t)^2 - 2 (g+1)S\cdot F} \left(\frac{e(g-2)}{2(P \cdot t)} F^\nu{}_\rho P^\rho + \frac{eg}{2} F^\nu{}_\rho t^\rho + \dot{t}^\nu - \frac{eg}{4 (P\cdot t)} S^{\rho\sigma}\partial_\nu F_{\rho\sigma}\right), \\
\dot{P}^\mu & = - e F^\mu{}_\nu \dot{X}^\nu - \frac{eg}{4} S^{\rho\sigma} \partial^\mu F_{\rho\sigma}, \\
\dot{S}^{\mu\nu} & = P^\mu \dot{X}^\nu - P^\nu \dot{X}^\mu + \frac{eg}{2} \left(S^{\mu}{}_\rho F^{\rho\nu} - S^\nu{}_\rho F^{\rho\mu}\right),
\end{align}
\end{subequations}

In the $g = 2$ case, we recover the equations of motion in \cite{OanceaK22}, considering that their equations are approximated up to first order in spin terms. Also, when the electromagnetic field vanishes, we recover \eqref{observer_xdot}.

Because the above equations of motion are rather complicated, and because we typically wish to study particles in the regime where $e S\cdot F/\cE^2 \ll 1$, one can consider the linearized equations,
\begin{subequations}
\begin{align}
\dot{X}^\mu & = P^\mu + \frac{S^\mu{}_\nu}{\cE^2} \left(\frac{e}{2}(g-2) F^\nu{}_\rho P^\rho + \frac{eg}{2}E \, F^\nu{}_\rho t^\rho + E  \, P^\rho \nabla_\rho t^\nu\right), \label{g_xdot} \\
\dot{P}^\mu & = - e F^\mu{}_\nu \dot{X}^\nu - \frac{eg}{4} S^{\rho\sigma} \partial^\mu F_{\rho\sigma}, \\
\dot{S}^{\mu\nu} & = P^\mu \dot{X}^\nu - P^\nu \dot{X}^\mu + \frac{eg}{2} \left(S^{\mu}{}_\rho F^{\rho\nu} - S^\nu{}_\rho F^{\rho\mu}\right).
\end{align}
\end{subequations}

\medskip

However, for our problem at hand, which is to find whether some motion is induced on a null hypersurface $\Sigma_v$, we can work directly with the exact equations \eqref{eom_gneq0}. For the particle to stay on $\Sigma_v$, its velocity must follow $\dot{X}^v = 0$, or equivalently $\dot{X}^\mu I_\mu = 0$, with $I = \partial_u$ \footnote{Since $I$ is the momentum with vanishing gyromagnetic ratio, see section \ref{s:chiral fermions}.}. However, since $I_\mu S^\mu{}_\nu = 0$, we immediately have, from \eqref{g_xdot} and \eqref{def p g},
\begin{equation}
\dot{X}^\mu I_\mu = - \frac{eg}{4} S\cdot F.
\end{equation}

Hence, we see that the condition for the massless chiral fermion to have motion induced on a null hypersurface is that, if we assume $S \cdot F \neq 0$, its gyromagnetic ratio $g$ must vanish. The only kind of massless chiral fermions with induced Carroll motions are those discussed in section \ref{s:chiral fermions}.

\section{Conclusions}

In this article, we have looked at the motion of elementary particles in a Lorentzian spacetime and determined whether these motions can be induced on null hypersurfaces, which are Carroll geometries. 

First, we studied usual spinless massless particles, and have recalled the known result that these particles induce motion on null hypersurfaces, however trivial: from the Carrollian point of view, these particles do not move. While trivial, this serves as the basis to explain the realization of Carroll theories from an ambient point of view.

While the motion of spinless massless particles is trivial, photons do have spin. We saw that such particles also have motion induced on null hypersurfaces, but this time the motion might be non trivial, due to Wigner-Souriau translations. The effect of these ``translations'' being that the trajectory of the particle is observer dependent, which leads to apparent motion on the Carroll structures. From the intrinsic Carrollian point of view, these non-trivial motions are somewhat artificial, in that they depend on an observer which lives in the bulk, not on the Carroll geometry.

Going up in complexity, we also investigated charged and massless particles with spin, also called massless chiral fermions. These particles are not elementary, but quasiparticles \cite{Venemaetal16}. Whether motion is induced on null hypersurfaces depends on the value of the gyromagnetic ratio $g$ of these particles. If $g = 0$, we find induced motions on a Carroll hypersurface. This motion is interesting for two reasons. First, because it matches a previously derived intrinsic Carrollian model found in \cite{MarsotZCH22}. Second, because it displays the same equations of motion as the Hall law (as already noted in \cite{MarsotZCH22}).

Lastly, if the gyromagnetic ratio of the particles does not vanish, $g \neq 0$, we find that it is impossible for the motion of the particle to stay on a null hypersurface. In other words, massless chiral fermions with a non-vanishing gyromagnetic ratio cannot induce Carroll motions.

\section*{Acknowledgments}
The author is grateful to Marius Oancea for enlightening discussions. The author also thanks Peter Horv\'athy and Pengming Zhang for comments.

\bibliographystyle{mystyle}
\bibliography{mybib}

\end{document}